# Autocompensating Quantum Cryptography

**Donald S. Bethune and William P. Risk**

IBM Almaden Research Center
650 Harry Road
San Jose, CA 95120-6099, USA
bethune@almaden.ibm.com and risk@almaden.ibm.com

**Abstract.** Quantum cryptographic key distribution (QKD) uses extremely faint light pulses to carry quantum information between two parties (Alice and Bob), allowing them to generate a shared, secret cryptographic key. Autocompensating QKD systems automatically and passively compensate for uncontrolled time-dependent variations of the optical fiber properties by coding the information as a differential phase between orthogonally-polarized components of a light pulse sent on a round trip through the fiber, reflected at mid-course using a Faraday mirror. We have built a prototype system based on standard telecom technology that achieves a privacy-amplified bit generation rate of ~1000 bits/s over a 10-km optical fiber link. Quantum cryptography is an example of an application that, by using quantum states of individual particles to represent information, accomplishes a practical task that is impossible using classical means.

## 1. Introduction

### 1.1. Quantum Cryptography

Quantum cryptography or quantum key distribution (QKD), invented by Bennett and Brassard and described in their classic "BB84" paper [1], uses the states of single photons to carry information between two parties, traditionally called Alice and Bob. By performing a joint, remote experiment with these photons, followed by a public discussion of the results, Alice and Bob can generate a shared cryptographic key. Furthermore, by exploiting the quantum nature of the photons, they can be sure that their key is secure, even though an eavesdropper (Eve) has unrestricted access to the photons as they travel between Alice and Bob through an optical fiber or free space, *and* has full knowledge of all other classical communications between them. Bennett and co-workers also provided the first experimental demonstration of QKD in 1989 [2], using weak light pulses transmitted 30 cm through free space. Subsequently, numerous groups have built a variety of systems for carrying out QKD, both over optical fiber and through free space. Ref. [3] provides an excellent recent review of the field.

In its simplest (and to date, most practical) form, quantum information is encoded in the states of single photons (or extremely faint coherent light pulses) that travel from Alice to Bob. An equivalent [4], but more difficult, method for conveying quantum information is to let Alice and Bob share an entangled pair of photons [3-8]. In that case, a definite value of a physical quantity (such as angular momentum or energy) is associated with the pair state, but *not* with either of the photons individually. For these entangled-pair states, very strong correlations will exist in the results of measurements Alice and Bob make on their separate photons, allowing them to generate a cryptographic key from these results and public discussion of them. Eavesdropping necessarily reduces the degree of correlation between Alice and Bob's measurements, so the degree of correlation in the data can be used to calculate an upper bound on the amount of information leaked to possible adversaries.

*1.2. Autocompensating interferometry as a basis for QKD*

A general challenge for quantum information is to choose a coding scheme that avoids disturbance by the dominant noise sources. For QKD based on either fiber or free space optical systems, fluctuations of the medium through which the photons travel are a source of noise. In free space, atmospheric density fluctuations are locally highly isotropic. Because polarization coding is essentially a differential phase encoding between two orthogonal polarization states, density fluctuations give rise to 'common mode' phase fluctuations, which do not degrade the contrast between the polarization basis states. Perturbations of the photon's propagation direction by density fluctuations only reduce the collection efficiency and key generation rate.

The situation is quite different when QKD is carried out over optical fibers. Optical fiber systems are a natural choice for QKD over distances up to several tens of kilometers, particularly since fiber optic networks already connect many computers. Unfortunately standard optical fiber (for example SMF-28) has small but significant birefringence due to variations in its structure and composition and due to mechanical stresses caused, for example, by bends and twists. Other optical components used in a fiber optic network can also introduce birefringence. These effects lead to random time-varying changes of polarization state, so that constant measurement of the properties of the optical fiber and active feedback control of some sort of compensating optical devices is required to use polarization as a basis for coding quantum information.

While this active approach has been successfully used in fiber-optic QKD systems [9], it is also possible to build a system that inherently compensates for changing fiber characteristics by using a differential coding scheme. This idea can be understood by considering Martinelli's observation that light sent on a round trip through an optical fiber terminated with a Faraday mirror returns in the polarization state orthogonal to that in which it started, independent of the optical properties of the fiber [10]. This property arises from the fact that a Faraday mirror exchanges orthogonal polarization states, with the consequence that the total phase accumulated by light over the course of a round-trip is *independent of the input polarization*. Because of this, information encoded as a differential phase between two orthogonal polarization states will be unaffected by birefringence in the system, provided that this birefringence does not change in the time it takes light to make a round-trip through the fiber ($2n/c \sim 10$ µs/km). The idea that a QKD system could be made insensitive to the state of the fiber by using round-trip propagation with a Faraday reflection mid-course was developed independently by groups at the University of Geneva [11-14] and IBM [15,16], who built systems that used 1.3-µm light. Operation at 1.55-µm, preferred for long-haul telecom systems, has also been demonstrated [17,18].

*1.3. Mathematical treatment of autocompensating system*

Martinelli originally used the Jones matrix formalism to show that a Faraday mirror can "orthoconjugate" light–i.e., generally transform any polarization state into an orthogonally-polarized state. This formalism can be extended to describe autocompensating QKD systems. We represent the transformation of the polarization state of light propagating forward and backward through an optical component or system by matrices $T$ and $T_R$ given by

$$T = V \cdot \begin{bmatrix} e^{iX} & 0 \\ 0 & e^{iY} \end{bmatrix} \cdot U \qquad T_R = U_R \cdot \begin{bmatrix} e^{iX} & 0 \\ 0 & e^{iY} \end{bmatrix} \cdot V_R \qquad (1)$$

Here $U$ and $V$ are arbitrary unitary matrices that transform between the eigenmode basis of the optical system and any desired bases at the system ends, and $e^{iX}$ and $e^{iY}$ describe the phase and amplitude changes for each of the eigenmodes. The matrices for backward propagation through such components are derived from their forward counterparts by transposing and negating the off-diagonal elements (following the convention that backward propagating fields be described in a right-handed coordinate system with reversed $z$- and $x$- axes) [19-21]. For reciprocal components, $X$ and $Y$ are independent of propagation direction. For non-reciprocal components such as Faraday rotators, reversing the propagation direction also changes the sign of the Faraday rotation angle, given by $\gamma = \Delta n_c k_o L = VBk_o L$ where $\Delta n_c = \frac{1}{2}(n_- - n_+)$ [with $n_\pm$ the refractive indices for positive and negative helicity light], $k_o$ is the light wavevector *in vacuo*, $V$ is the Verdet constant, $B$ is the component of magnetic field parallel to $k_o$, and $L$ is the path length in the medium. Thus the forward and backward polarization transformations through a Faraday rotator are related by $F(\gamma) = F_R(-\gamma)$, and can be written as $2 \times 2$ rotation matrices for angles $\pm \gamma$. Purely optically-active media are described similarly, but with no sign change for $\gamma$ in the reverse direction.

A Faraday mirror, which is a combination of a $45°$ Faraday rotator and a normal mirror, transforms polarization states according to the Jones matrix:

$$FM = \frac{1}{2}\begin{bmatrix} 1 & 1 \\ -1 & 1 \end{bmatrix} \cdot \begin{bmatrix} 1 & 0 \\ 0 & -1 \end{bmatrix} \cdot \begin{bmatrix} 1 & -1 \\ 1 & 1 \end{bmatrix} = \begin{bmatrix} 0 & -1 \\ -1 & 0 \end{bmatrix} \qquad (2)$$

$FM$ can be multiplied by a factor $r$ to account for attenuation and phase shift due to the optics. It is readily verified that for unitary matrices such as $U$, $U_R \cdot FM \cdot U = FM$. Thus, for a round trip through a system described by $T$ and terminated by a Faraday mirror, the overall Jones matrix $M$ is given by $M = T_R \cdot FM \cdot T$, or

$$M = T_R \cdot \begin{bmatrix} 0 & -1 \\ -1 & 0 \end{bmatrix} \cdot T = e^{i(X+Y)} \cdot r \cdot \begin{bmatrix} 0 & -1 \\ -1 & 0 \end{bmatrix} = e^{i(X+Y)} \cdot r \cdot FM \quad . \tag{3}$$

$M$ is simply given by the Faraday mirror matrix multiplied by overall phase and amplitude attenuation factors. If several components described by $T_i$ of the form in Equation (1) are concatenated before Faraday reflection, a round trip is described by the matrix $M$ with $X$ and $Y$ replaced by $\sum X_i$ and $\sum Y_i$, respectively. The final polarization state of light after a round trip is still orthogonal to the initial state, independent of the properties of the optical system described by the $T_i$, and the total phase accumulated is $\Sigma(X_i + Y_i)$, independent of the initial polarization state.

When a light pulse with horizontal polarization state $\psi_i = \begin{bmatrix} a_H \\ 0 \end{bmatrix}$ is presented at the input of such a system, the state that is returned is

$$\psi_r = e^{i(\Sigma X_i + \Sigma Y_i)} \cdot r \cdot \begin{bmatrix} 0 & -1 \\ -1 & 0 \end{bmatrix} \psi_i = -e^{i(\Sigma X_i + \Sigma Y_i)} \cdot r \cdot \begin{bmatrix} 0 \\ a_H \end{bmatrix} \quad , \tag{4}$$

which has vertical linear polarization. Similarly, when the vertically polarized state $\begin{bmatrix} 0 \\ a_V \end{bmatrix}$ is input to the system, the returned state is $-e^{i(\Sigma X_i + \Sigma Y_i)} \cdot r \cdot \begin{bmatrix} a_V \\ 0 \end{bmatrix}$, and is horizontally polarized. The prefactors are the same in both cases; thus no differential phase shift or attenuation is introduced between two orthogonally polarized light pulses that travel on a round trip through the same sequence of optical elements terminated by a Faraday mirror.

For QKD, this result allows Alice and Bob to robustly encode quantum information in the differential phase between orthogonally polarized pulses. In the preceding discussion, we assumed that the modal phase shifts $X$ and $Y$ were the same for outgoing and returning light. However, this will only be true if the system optical properties vary slowly compared to the round trip time for light. By deliberately violating this condition using fast phase modulators, Alice and Bob can introduce useful differential phase shifts between the orthogonal polarization states. These shifts can be read out at Bob's station either polarimetrically (by combining the orthogonally polarized waves) or interferometrically (by combining the waves with parallel polarization) using appropriate optics. Systems built at IBM based on these two read-out methods are described below in Sections 2 and 3, respectively.

## 2. Experiments using bulk optical components

*2.1. Experimental setup*

Figure 1 shows the initial version of our quantum cryptography set up. Alice and Bob's setups, in the same lab but with separate computers and independent equipment, are only connected via a 10-km (or 20-km) SMF-28 optical fiber link that carries both quantum information at 1.3-μm and wavelength multiplexed bi-directional timing and coordination pulses using 1.55-μm transceivers. They also communicate via the building LAN to carry on public discussions. In this optical design, a horizontally polarized pulse from the 1.31 μm DFB laser (Lucent D2304G) passes through a variable attenuator and bat-wing polarization adjuster (not shown) and then is fiber-coupled to a bulk optical polarization and analysis module. The pulse passes through two polarizing beamsplitters PBS1 and PBS2 (with canceling ±45° rotations between due to the Faraday rotator and WP1), and then after a 45° rotation by WP2, is split by PBS3 into *H*- and *V*- polarized components with amplitudes we can describe by the vector

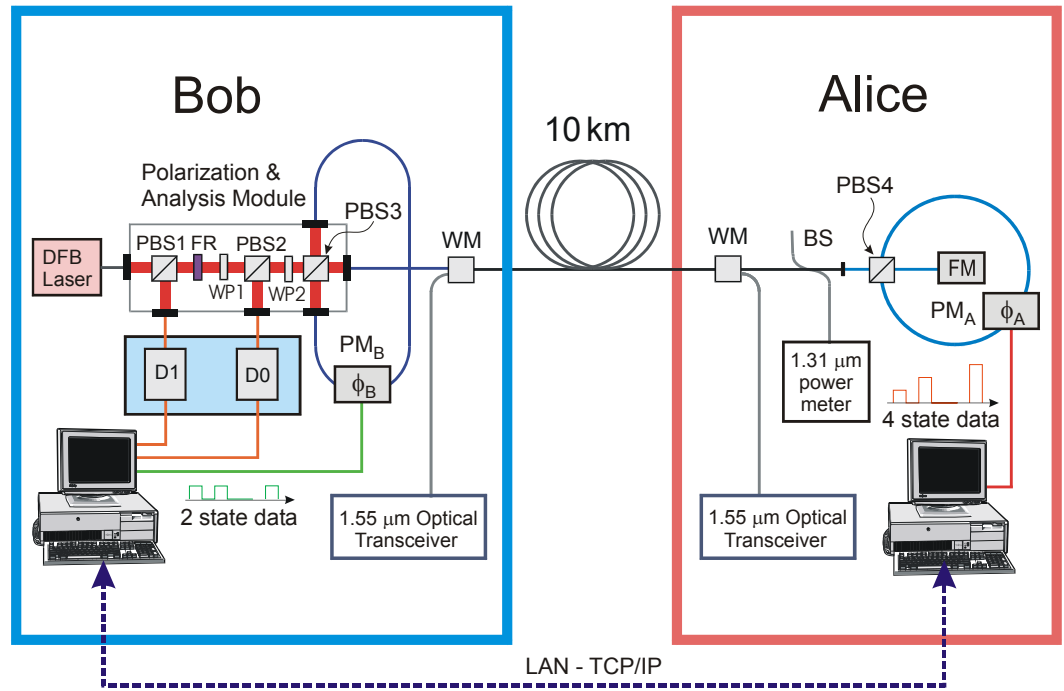

**Figure 1.** QKD system using free-space bulk optical polarizing beamsplitters (PBS). FR – Faraday Rotator; WP1, WP2 - λ/2 waveplates; WM – 1310/1550 nm multiplexers; $PM_{A,B}$ – phase modulators; FM – Faraday mirror; $D_{0,1}$ - detectors; DFB Laser – distributed feedback 1310 nm laser, pulsed to give ~50 ps pulses.

$$\psi_0 = \begin{bmatrix} a_H \\ a_V \end{bmatrix} = \begin{bmatrix} a_H \\ 0 \end{bmatrix} + \begin{bmatrix} 0 \\ a_V \end{bmatrix} = \psi_{0H} + \psi_{0V} \quad .$$

The *H*-component (*H*) is launched into the SMF-28 fiber, while the *V*-component (*V*) is delayed a time τ ~30 ns by a loop of polarization-maintaining fiber before a second encounter with PBS3 directs it into the fiber toward Alice. Since pulses *H* and *V* pass through the system at different times, they can be independently traced through the system in order to determine their states upon returning to PBS3.

### 2.1.1. The H-polarized pulse

After passage through the fiber, *H* arrives at PBS4 in an arbitrary elliptical polarization state given by $\psi_{1H} = T\psi_{0H}$, where *T* describes propagation between PBS3 and PBS4. PBS4 separates *H* into *x*- and *y*-components corresponding to the axes of the polarization maintaining fibers connecting PBS4 to Alice's phase modulator (PM$_A$) and the Faraday mirror. By placing PM$_A$ in a loop as shown in figure **2**, an annealed-proton-exchange (APE) LiNbO$_3$ waveguide modulator can be used to apply an overall phase shift to the arbitrarily polarized pulse, even though the waveguide transmits only one linear polarization (*y*). The *y*-polarized component of $\psi_{1H}$ is sent by PBS4 directly to PM$_A$. The *x* component travels to the Faraday mirror, is rotated to *y*-polarization by the FM, returns to PBS4, and is coupled into the loop by PBS4. PM$_A$ is offset from the loop midpoint by an optical path *L* equal to the path from PBS4 to the FM, so that the counter-propagating *x* and *y* pulse components meet in PM$_A$ having each traveled the same distance (*C*/2+*L*) from PBS4, where *C* is the loop circumference. This arrangement places the modulator at the exact midpoint of the overall optical path. As the *x* and *y* components pass through PM$_A$ at the same time in opposite directions, an equal phase shift $\phi_{AH}$ is imparted to both of them. The component that was originally *x*-polarized continues counterclockwise around the loop to PBS4. The component that was originally *y*-polarized travels clockwise around the loop to PBS4 and then onto the FM, where it is rotated to *x*-polarization and sent back to PBS4, where the two components are reunited.

The overall combined effect of PBS4, the fiber loop and phase modulator, the Faraday mirror, and the attenuation due to WM and Alice's other components (which is included in *r*), can be described by the matrix:

$$FM' = r \cdot e^{i\phi_s} \cdot e^{i\phi_{AH}} \cdot \begin{bmatrix} 0 & -1 \\ -1 & 0 \end{bmatrix} . \qquad (5)$$

Here $\phi_s = k_x L + k_y (C+L)$ is a static phase shift that takes into account the birefringence of the polarization-maintaining fiber path. This arrangement is thus equivalent to a standard Faraday mirror plus a controllable phase shift, $\phi_{AH}$, that can be intentionally applied using Alice's phase modulator. After reflection, phase shifting and attenuation, as represented by matrix FM', pulse *H* returns through the fiber to PBS3, arriving in the state

$$\psi_{2H} = T_R \cdot FM' \cdot T \cdot \psi_{0H} = r \cdot e^{i(X+Y)} \cdot e^{i\phi_s} \cdot e^{i\phi_{AH}} \cdot \begin{bmatrix} 0 & -1 \\ -1 & 0 \end{bmatrix} \cdot \begin{bmatrix} a_H \\ 0 \end{bmatrix}$$
$$= -r \cdot e^{i(X+Y)} \cdot e^{i\phi_s} \cdot e^{i\phi_{AH}} \cdot \begin{bmatrix} 0 \\ a_H \end{bmatrix} \tag{6}$$

Thus $H$, which originally was horizontally-polarized, returns to PBS3 vertically-polarized and is directed through Bob's delay loop. The loop adds both a fixed phase, $\phi_\tau$, and a controllable phase $\phi_{BH}$ that Bob can apply using PM$_B$. $H$ then arrives at PBS3 for the second time in the state

$$\psi'_{2H} = -r \cdot e^{i(X+Y)} \cdot e^{i\phi_s} \cdot e^{i\phi_\tau} \cdot e^{i\phi_{AH}} \cdot e^{i\phi_{BH}} \cdot \begin{bmatrix} 0 \\ a_H \end{bmatrix} = -r' \cdot e^{i(\phi_{AH}+\phi_{BH})} \cdot \begin{bmatrix} 0 \\ a_H \end{bmatrix}, \tag{7}$$

where the attenuation and fixed phase factors have been combined into the new constant $r'$.

### 2.1.2. *The V-polarized pulse*

The vertically polarized pulse $V$, initially in state $\psi_{0V}$, travels through the system in an analogous way, but passes through the delay loop immediately on departure. As it did for $H$, the loop adds the fixed phase $\phi_\tau$ and a phase shift $\phi_{BV}$, controlled by PM$_B$. $V$ returns to PBS3 in the state

$$\psi'_{2V} = T_R \cdot FM' \cdot T \cdot \psi_{1V} = r \cdot e^{i(X+Y)} \cdot e^{i\phi_s} \cdot e^{i\phi_\tau} \cdot e^{i\phi_{AV}} \cdot e^{i\phi_{BV}} \cdot \begin{bmatrix} 0 & -1 \\ -1 & 0 \end{bmatrix} \cdot \begin{bmatrix} 0 \\ a_V \end{bmatrix}$$
$$= -r' \cdot e^{i(\phi_{AV}+\phi_{BV})} \cdot \begin{bmatrix} a_V \\ 0 \end{bmatrix}, \tag{8}$$

which is horizontally-polarized and has an intentional phase shift $\phi_{AV} + \phi_{BV}$.

Recall that the first pulse, $H$, is directed through Bob's delay loop on its return to PBS3 and arrives at PBS3 for the second time after a delay $\tau$. Since $\tau$ is also the delay between $H$ and $V$, $H$'s second arrival at PBS3 coincides with $V$'s arrival. The two orthogonally-polarized pulses thus emerge from PBS3 at the same time, overlapping in space, and travel together toward WP2. The polarization state of the recombined pulse as it leaves PBS3 can be expressed by the vector:

$$\psi_3 = -r' \cdot \begin{bmatrix} e^{i(\phi_{AV}+\phi_{BV})} a_V \\ e^{i(\phi_{AH}+\phi_{BH})} a_H \end{bmatrix} = -r' \cdot e^{i\Sigma\phi} \begin{bmatrix} e^{i\Delta\phi} a_V \\ e^{-i\Delta\phi} a_H \end{bmatrix}, \tag{9}$$

where $\Sigma\phi = \dfrac{(\phi_{AV}+\phi_{BV})+(\phi_{AH}+\phi_{BH})}{2}$ and $\Delta\phi = \dfrac{(\phi_{AV}+\phi_{BV})-(\phi_{AH}+\phi_{BH})}{2}$.

In typical operation, Alice sets $\phi_{AH} = 0$ while the first pulse passes through the modulator and then rapidly switches to $\phi_{AV} \neq 0$ in a time short compared to the delay $\tau$ between the $H$ and $V$ pulses. Similarly, Bob turns his modulator off for all pulses leaving his station ($\phi_{BV} = 0$), but sometimes applies a phase $\phi_{BH} \neq 0$ to returning pulses, so that $\Delta\phi = \dfrac{\phi_{AV} - \phi_{BH}}{2}$.

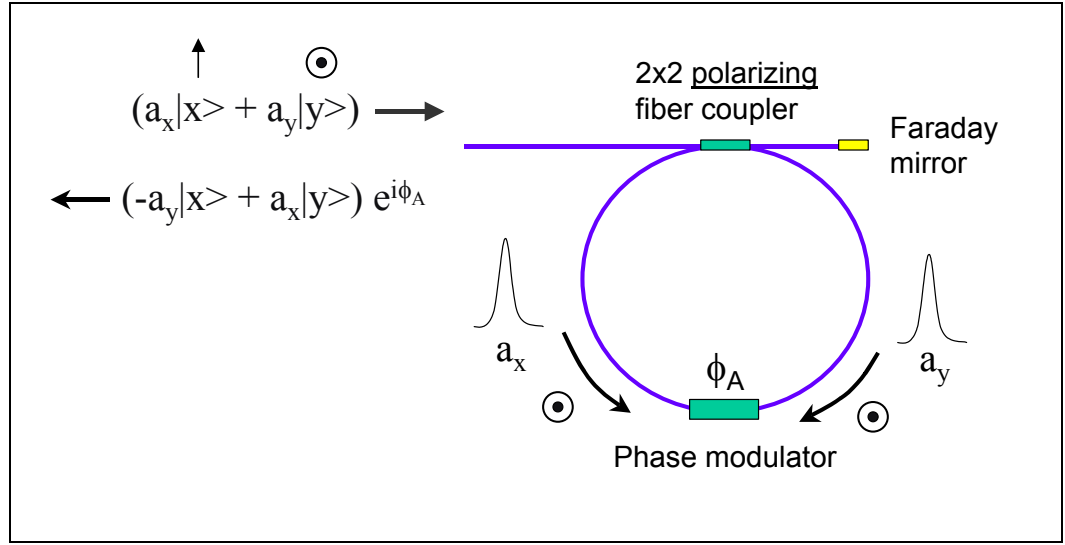

**Figure 2.** An arrangement for Faraday reflecting an arbitrarily polarized pulse with an overall phase shift $\phi_A$ using a single-polarization phase modulator. [Click this link to see an animated depiction of the operation of this arrangement]

### 2.1.3. *Polarization analysis for decoding the quantum bits*

After leaving PBS3, the light represented by state $\psi_3$ is rotated by WP2 to the final state

$$\psi_4 = -i \cdot r' \cdot e^{i\Sigma\phi} \cdot \begin{bmatrix} -\cos 2\theta & \sin 2\theta \\ \sin 2\theta & \cos 2\theta \end{bmatrix} \begin{bmatrix} e^{i\Delta\phi} a_V \\ e^{-i\Delta\phi} a_H \end{bmatrix} . \qquad (10)$$

Recalling that the state $\psi_0 = \begin{bmatrix} a_H \\ a_V \end{bmatrix}$ was originally produced by the passage of a purely horizontally-polarized state through WP2 in the forward direction, we can write:

$$\psi_0 = \begin{bmatrix} a_H \\ a_V \end{bmatrix} = -i \cdot \begin{bmatrix} \cos 2\theta & \sin 2\theta \\ \sin 2\theta & -\cos 2\theta \end{bmatrix} \cdot \begin{bmatrix} 1 \\ 0 \end{bmatrix} = -i \cdot \begin{bmatrix} \cos 2\theta \\ \sin 2\theta \end{bmatrix} , \qquad (11)$$

so that

$$\psi_4 = -r' \cdot e^{i\Sigma\phi} \cdot \begin{bmatrix} -\cos 2\theta & \sin 2\theta \\ \sin 2\theta & \cos 2\theta \end{bmatrix} \begin{bmatrix} e^{i\Delta\phi} \sin 2\theta \\ e^{-i\Delta\phi} \cos 2\theta \end{bmatrix}$$
$$= -r' \cdot e^{i\Sigma\phi} \cdot \begin{bmatrix} -i\sin\Delta\phi \sin 4\theta \\ \cos\Delta\phi - i\sin\Delta\phi\cos 4\theta \end{bmatrix} \quad . \tag{12}$$

In the ideal case WP2 is rotated to the angle $\theta = \pi/8$, and thus

$$\psi_4 = -r' \cdot e^{i\Sigma\phi} \cdot \begin{bmatrix} -i\sin\Delta\phi \\ \cos\Delta\phi \end{bmatrix} \quad . \tag{13}$$

This shows that PBS2 directs all of the light to $D_0$ for $\Delta\phi$ an even multiple of $\pi/2$, and to $D_1$ (via PBS1) for $\Delta\phi$ an odd multiple of $\pi/2$. If WP2 is rotated by $\delta$ away from its ideal angle, the fraction of the intensity leaking to $D_1$ has a minimum value $(4\delta)^2$ (when $\Delta\phi = \pi$), while the minimum intensity to $D_0$ (for $\Delta\phi = 0$) will be 0, independent of $\delta$. A 2° misalignment of WP2, for example, will misroute ~2% of the intensity for even $\Delta\phi/(\pi/2)$, while for odd $\Delta\phi/(\pi/2)$ all photons will be routed correctly. Thus, on average, a 2° misalignment of $\theta$ will contribute ~1% to the bit error rate (BER).

With their shared control over the differential phase shift $\Delta\phi$ in Equation (13), Alice and Bob can implement the 4-state BB84 protocol. A possible assignment of basis and bit values to Alice and Bob's phase shifts is given in Table **1**. When Alice and Bob match bases, with both choosing either an odd or an even multiple of $\pi/2$, the photons are deterministically routed to a specific detector, depending on Alice's choice of bit.

| Basis, Bit | | odd, 0 | even, 1 | odd, 1 | even, 0 |
|---|---|---|---|---|---|
| Basis | $\phi_B \downarrow / \phi_A \rightarrow$ | $-\pi/2$ | $0$ | $\pi/2$ | $\pi$ |
| odd | $-\pi/2$ | $D_0$ | ? | $D_1$ | ? |
| even | $0$ | ? | $D_0$ | ? | $D_1$ |

**Table 1.** Example of an assignment of bit and basis to various $\phi_A$, $\phi_B$ combinations. For deterministic cases the detector receiving the photon is indicated, while random arrival cases are indicated by **?**.

*2.2. Single-photon detection using pulse biased InGaAs avalanche photodiode (APD) detectors*

Quantum cryptography requires detection of single photons. The circuit used for this task is shown in figure **3a**. Two Fujitsu FPD5W1KS InGaAs detectors, optimally cooled to ~118 K and DC biased just below their reverse breakdown voltage (~30 V), are biased above breakdown for ~ 1.5 ns by a 3.3 V pulse at times when photons are expected to arrive. To avoid having the photocurrent pulse buried in the capacitive transient induced on the output line by the bias pulse (see, for example, the beginning of the scope traces in figure **3b**), the circuit shown in figure **3a** implements a simple transient canceling scheme. SMA tees are connected to the APD cathode and anode, with 50-Ω matched transmission lines and pulse-bias input and signal-output lines attached as shown. The open upper delay line reflects the bias pulse without inversion, while the shorted lower delay line gives an inverted reflection of the capacitive transient. The reflected bias pulse drives a second transient, which adds to and cancels the returning inverted transient, thus holding the anode voltage constant. The photocurrent pulse due to a detected photon, which coincides with only one of the bias pulses, is not cancelled, and gives a signal above a nearly flat baseline, as shown by the middle trace of figure **3b**. This signal pulse is amplified and sent to an electronic gate (described in detail in [16]), giving the final output pulse shown by the lower trace of figure 3b, overlaid on a trace obtained with no photon signal present. The output pulses are sent to a discriminator followed by appropriate logic and counting modules. The transient cancellation is independent of pulse amplitude, pulse shape, and photodiode characteristics.

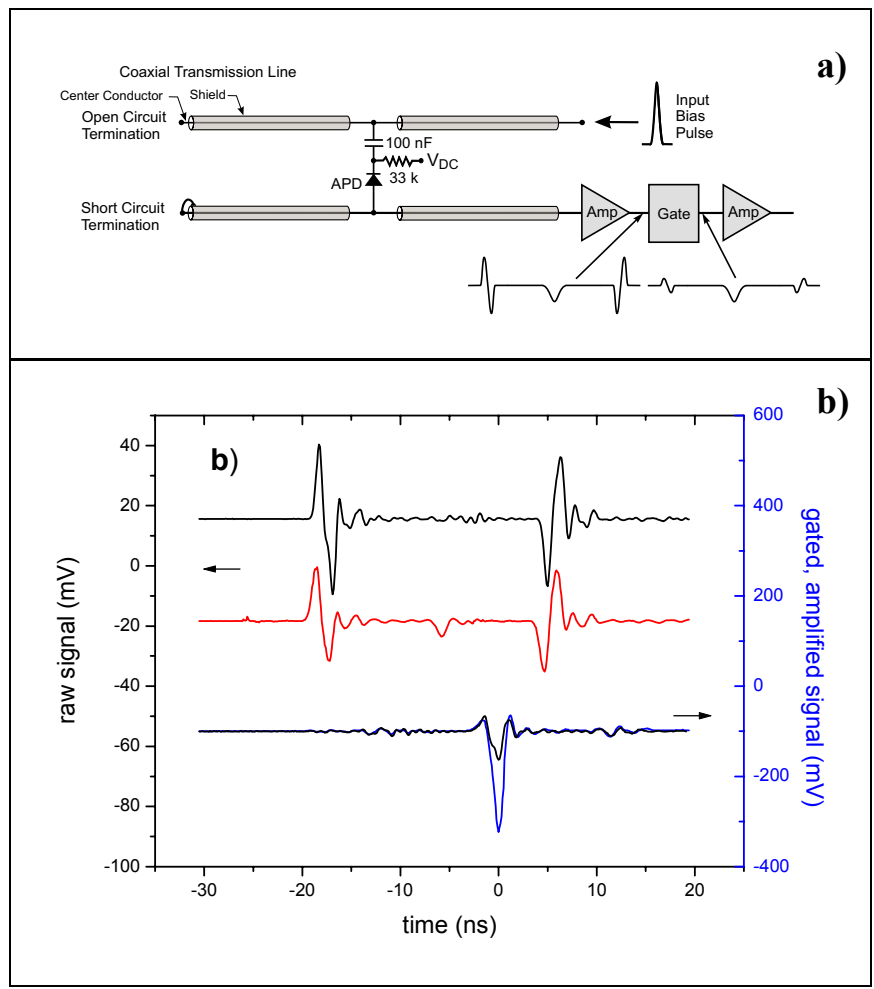

**Figure 3. a)** Detector circuit diagram. [Click on this link to see an animated depiction of the operation of this circuit] **b)** Upper trace: raw detector output with no detected photon; middle trace: raw output with detected photon; lower trace: signal outputs with and without detected photon after amplification and gating.

The detectors are pulse biased at 1 MHz, and with peak voltages ~2 V above breakdown a quantum efficiency $\eta \sim 20\%$ with dark count probability $\sim 2\text{–}4\times10^{-5}$ is achieved. The 1.55-µm timing pulses sent from Bob to Alice are used to synchronize the modulator $PM_A$ to the arriving light pulses, and are echoed from Alice to Bob to trigger Bob's detector bias pulser and drive a counter to keep track of the pulse number. The Fujitsu detectors have worked well, but operate best at quite low temperature and unfortunately are no longer available. However, recent work has identified devices (JDS Uniphase EPM 239 AA SS) that have acceptably good performance even above 200 K, making thermoelectric cooling an attractive option [22-24].

## 3. All fiber-optical setup

### 3.1. Experimental set-up

An improved version of the experimental apparatus is shown in figure 4. This version incorporates an all-fiber version of Bob's delay loop and pulse analysis optics that follows the elegant design of Ribordy et al. [14], with the slight modification that polarization maintaining fiber is used throughout, so no polarization adjusters are needed in the phase-sensitive portion of the optical path. Bob, rather than initially splitting a 45° polarized beam using a bulk-optical polarizing beam-splitter cube, splits his linearly-polarized input pulse into two replicas using a 2×2 fiber coupler constructed with polarization-maintaining fiber. The upper leg of the splitter is connected to a fiber-optic polarizing beam splitter (PBS) with a connector modified to couple two polarization-maintaining fibers with their fast axes perpendicular. The delayed pulse from the lower branch of the coupler reaches the PBS without polarization rotation, and leaves Bob's station polarized orthogonally to the undelayed pulse, as in the earlier setup.

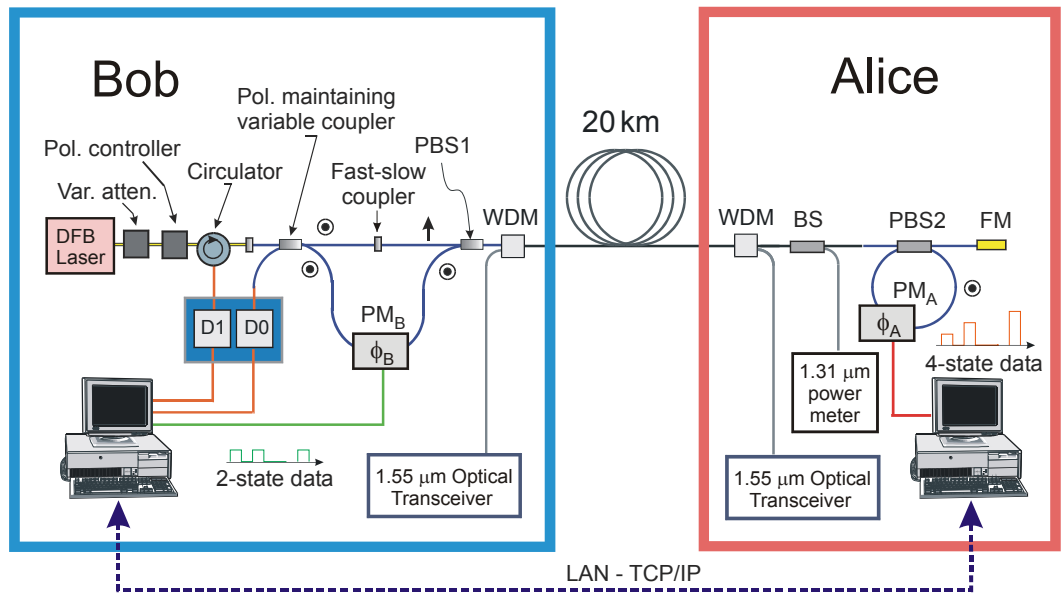

**Figure 4.** Experimental arrangement for all fiber-optical QKD system.

Alice also incorporates a fiber PBS with polarization-maintaining fiber to implement her phase modulator loop. Keeping the light in single mode fiber at both stations reduces the optical loss significantly, and most importantly, gives on/off interferometric switching contrast at the variable coupler of ~ 650:1, an order-of-magnitude higher than the 62:1 contrast obtained with the bulk-optic polarizers. It is also simpler to use a fiber optic circulator module to direct light to $D_1$ than to use the waveplate / Faraday-rotator / bulk PBS combination shown in figure **1**.

If we assume the initial coupler is symmetric, with amplitude transmission and reflection coefficients $t$ and $i \cdot r$ respectively ($r$ and $t$ real), we obtain expressions for the final amplitudes in the $D_0$ and $D_1$ channels analogous to those in Eqn. 12, with the substitutions $\sin(4\theta) \to 2i \cdot r \cdot t$ and $\cos(4\theta) \to (t^2 - r^2)$. The minimum probability for photons to be routed to $D_1$ is $(t^2-r^2)^2$. A non-ideal coupler with a 55:45 power splitting ratio, for example, will direct ~ 1% of the photons to $D_1$ at nominal null ($\Delta\phi = \pi$). With careful adjustment of the polarization maintaining variable coupler, we achieve a leakage rate 3–4 times lower than this.

## 4. Key Generation

### 4.1. System operation

Computers at Alice and Bob's stations, running LabVIEW™ programs, control the QKD system. They sequentially carry out the quantum information transfer, error correction, and privacy amplification. The weak 1.3-μm pulses and the 1.55-μm timing pulses are carried over a single SMF-28 optical fiber, while coordinating signals and data for error correction and privacy amplification are sent using TCP/IP over the building LAN. Data obtained for 10- and 20-km fiber links with the all-fiber system are shown in figure **5**. The raw key rates (squares) for photons detected with matching bases for Alice and Bob (~ ½ the total detection rate) are given by $\frac{1}{2}\mu\eta T f$, where $\mu$ is the mean number of photons/pulse leaving Alice's station, $\eta = 0.2$ is the detector quantum efficiency, $T = 0.08$ (0.032) is the 10 km (20 km) link transmittance (~ 8 dB loss for Bob's components with the remainder due to fiber loss) and $f = 1\,\text{MHz}$ is the pulse repetition frequency.

The error-corrected rates (circles) are based on the bits remaining after Alice and Bob's computers carry out error correction over the LAN. Starting with the raw key, they shuffle and block the key data 3 times, each time comparing row parities and bisectively searching out the errors in rows with parities that mismatch. The block sizes are chosen to leave ≲10 errors after three stages. The sparse remaining errors are found by comparing parities of randomly selected matching subsets of one-half of the remaining bits. If for a given draw the parities disagree (as will occur with probability $\frac{1}{2}$ if there are *any* remaining errors), a bisective search is used to find and eliminate the error. Success is assumed after 20 consecutive parity matches occur. In both the block and random subset operations, a bit is discarded for each parity bit revealed.

As a by-product of the error correction procedure, an estimate of the initial BER is obtained. This estimate is needed for privacy amplification. The actual BER's as a function of $\mu$ are also plotted (diamonds) in figure 5. Based on the estimated BER and the measured value of $\mu$, Alice and Bob compute an estimated upper bound, $E$, for the number of bits of information Eve possesses about their $N_{ec}$ error-corrected bits. To generate privacy-amplified key bits, Alice and Bob's computers continue to compute parities for matched subsets of one-half of the error-corrected bits (selected using a publicly transmitted random matrix), but now, rather than publicly comparing them, Alice and Bob keep these parities as secret key bits. $(N_{ec} - E - s)$ bits are generated in this way. The privacy amplification theorem asserts that, for a chosen $s$, Eve's information about the final key will be at most $\sim 2^{-s}/\ln 2$ bits [1,2].

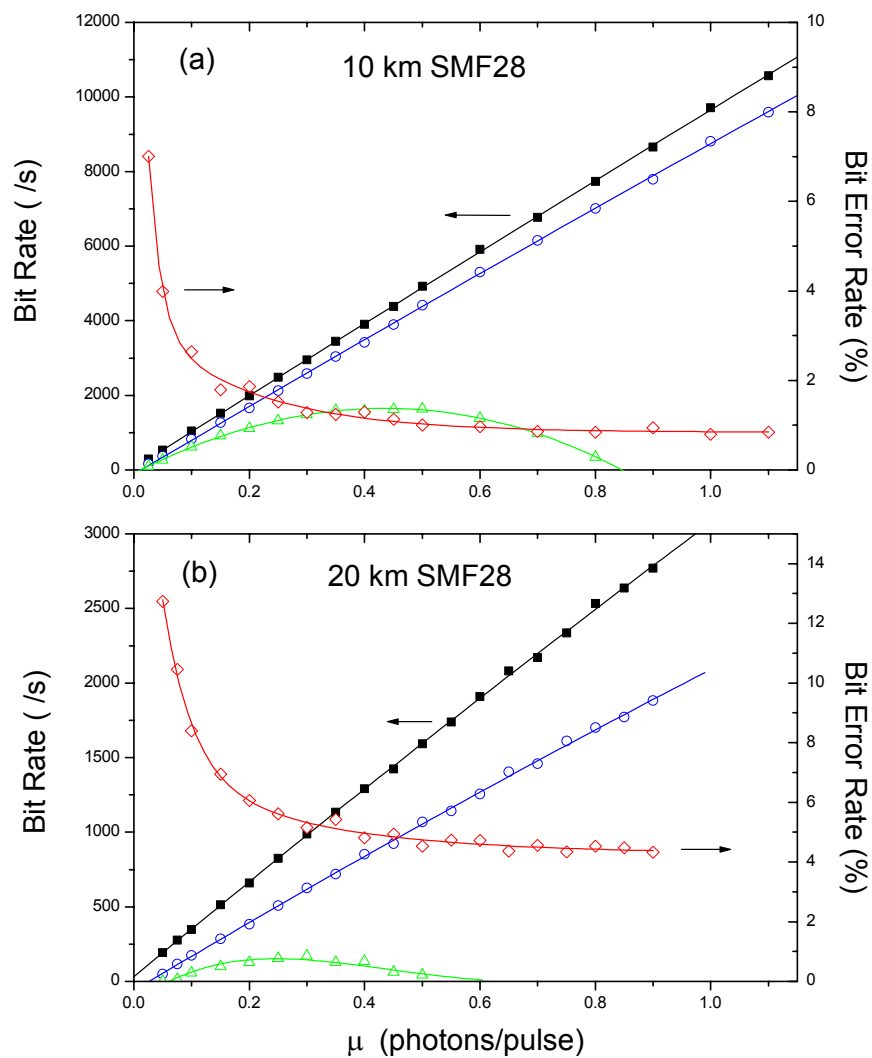

**Figure 5.** Key generation with the all-fiber system over **(a)** 10 and **(b)** 20 km links. The raw- (squares), error corrected- (circles) and privacy-amplified (triangles) key rates and bit error rates (diamonds) are plotted against $\mu$.

*4.2. Privacy amplification*

The maxima in the privacy amplified key rates versus $\mu$ seen in figure 5 reflect the tradeoff between low photon arrival rate and high BER at low $\mu$, and Eve's increasing ability to gain information about the pulses at high $\mu$. The degree of vulnerability of weak laser pulses to eavesdropping is a central and interesting question. The rates for privacy-amplified key generation in figure 5 were calculated using the simple BB84 estimate that Eve obtains fractions $2 \cdot \text{BER}$ and $\mu$ of the sifted bits using read-and-replace and beamsplitting attacks, respectively [1,2]. For a low BER this approximation gives a key rate that goes as $\sim \mu(1-\mu)$, with a maximum near $\mu = 0.5$ and falling to zero as $\mu \to 1$. For the 10-km link, using the BB84 leak estimate gives a maximum key rate ~1.5 kbit/s for $\mu = 0.3$, while for the 20-km link, the ~ 4 dB greater attenuation, higher backscattering, and consequently increased BER reduce the privacy-amplified key rate at $\mu = 0.3$ by about an order of magnitude, to 200 bit/s.

Analyses that consider more sophisticated eavesdropping attacks have been recently made [25-33], in some cases granting Eve powers within the realm of known physics, but far beyond those provided by today's technology. Lütkenhaus [29-31], for example, considers a scenario where Eve can measure the photon number for each pulse, block the single photon pulses, enhance the transmission of multi-photon pulses, and learn the exact state for pulses containing two or more photons, allowing perfect re-sending. Gilbert and Hamrick (G&H) [32] make a similar analysis, but allow Eve less information for 2-photon pulses, arguing that their state cannot be analyzed perfectly and that only those pulses where Eve and Bob each detect one photon contribute to Eve's knowledge of the error-corrected key. This is an important distinction, since double-photon pulses are relatively numerous.

Figure 6 compares the key generation rates versus $\mu$ calculated according to the prescriptions of these authors for various levels of system efficiency with those given by the simple BB84 estimate (shown in blue) [1,2]. The calculated curves assume dark count rates and backscattering levels typical of our system with the 10-km fiber link. The predicted net key rates are similar for ideal channels (quantum efficiency $\eta$ and link transmission $T$ both equal to 1), but diverge for lower $\eta$ and $T$. Using the prescription of G&H gives a bit rate of 200 bit/s for our 10-km system rather than the 1.5 kbit/s obtained with the simple BB84 prescription, and no net bits with for the 20-km link. Our 20-km BER is increased significantly (by 2-3%) owing to the greater intensity of backscattered light at 1.3-μm arriving at the detectors. One solution to this problem is to run Bob's laser intermittently, storing a train of pulses behind Alice's attenuator in a delay line [14], and detecting the returning photons during periods when no backscattered photons are present. While the system duty-factor is somewhat reduced using this approach, the very high cost of increased BER to key generation makes the trade-off highly worthwhile for longer distances.

In considering what level of data sacrifice is required for adequate privacy amplification, it is helpful to consider the point of view of the Geneva group [3,33]: whereas perfect security would be infinitely costly and impractical, practical security to a very high level is achievable with current level QKD systems if we accept quite reasonable limits on Eve's abilities. For example, large high-speed quantum memories with very long retention times are unavailable, and could be defeated simply by delaying sufficiently the exchange of basis information. Alternatively, Alice and Bob could generate all matching basis choices on the fly by using some initially shared secure key information, and never need to reveal the bases [34]! In either of these cases Eve would be forced to analyze the photons split from 2-photon pulses *without* a priori knowledge of the bases, which halves the information she can obtain about them. The optimum value of $\mu$ and the net rate of key generation for our 10-km link approximately double if the G&H leak estimate is modified in this way, as shown by one example in figure 6.

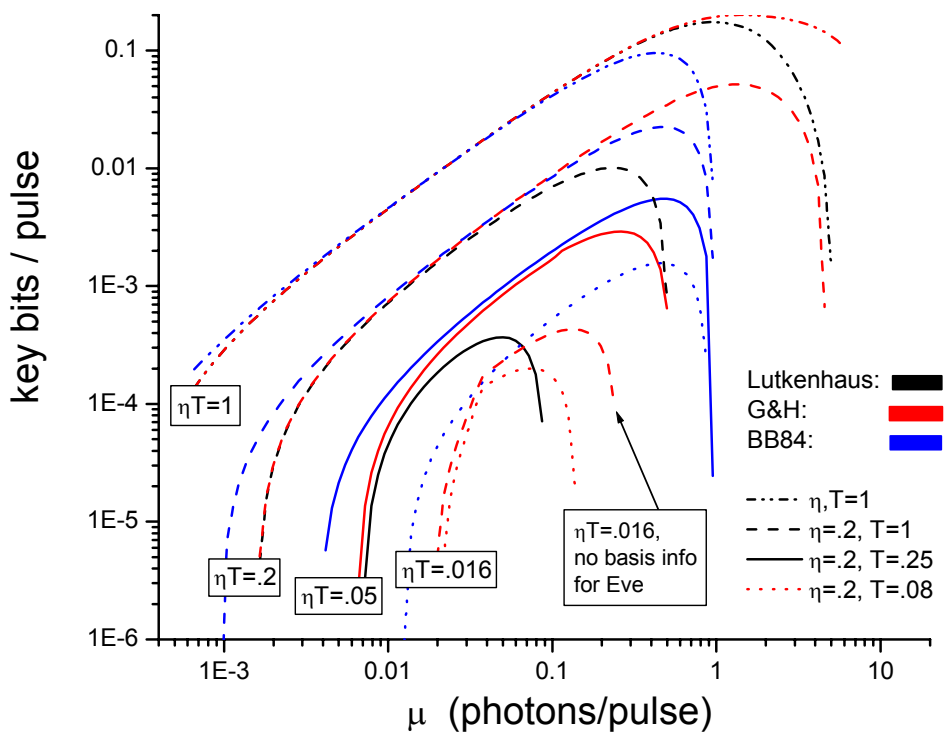

**Figure 6.** Secure key bits generated per pulse for several detector quantum efficiencies, $\eta$, and Alice-to-detector transmission values, $T$, according to the information leakage bounds given in [1,2], [30] and [32].

## 5. Conclusion

Autocompensating quantum cryptography systems make it possible to generate shared, secret cryptographic key data at kilobit-per-second rates over distances up to a few tens of kilometers. The advantage of the autocompensating approach is that it allows accurate optical readout of the quantum information with no monitoring or active control of the optical properties of the system, despite the presence of random, time-varying disturbances of the fiber link.   This is accomplished by using a differential phase coding that is immune to these disturbances.

## Acknowledgments

We gratefully acknowledge the support and inspiration for this work provided by Nabil Amer, helpful discussions with Charles Bennett, David DiVincenzo, Norbert Lütkenhaus and Hoi-Kwong Lo about security issues, and discussions about experimental issues with Paul Townsend and members of the GAP-Optique of the University of Geneva.